\documentclass[showpacs,preprintnumbers,superscriptaddress,amsmath,amssymb,twocolumn,nobibnotes,longbibliography,nofootinbib]{revtex4-2}
\usepackage[english]{babel}
\usepackage{mathtools}
\usepackage{graphicx}
\usepackage{mathrsfs}
\usepackage{amssymb}
\usepackage{xcolor}
\usepackage[colorlinks=true,citecolor=blue,urlcolor=blue]{hyperref}
\usepackage{amsmath}
\usepackage{bm}
\usepackage{physics}

\usepackage{nicefrac} 			

\begin{document}

\title{  Terahertz field-driven nonlinear Hall effect \\and other second order transport phenomena in two-dimensional tellurene	}

\author{M. D. Moldavskaya}
\affiliation{Physics Department, University of Regensburg, 93040 Regensburg, Germany}

\author{L. E. Golub}
\affiliation{Institute of Theoretical Physics and Halle-Berlin-Regensburg Cluster of Excellence CCE, University of Regensburg, 93040 Regensburg, Germany}

\author{E. M\"onch}
\affiliation{Physics Department, University of Regensburg, 93040 Regensburg, Germany}

\author{Chang Niu}
\affiliation{Elmore Family School of Electrical and	Computer Engineering, Purdue University, West Lafayette, Indiana 47907, United States}
\affiliation{Birck Nanotechnology Center, Purdue University, West Lafayette, Indiana 47907, United
	States}

\author{Peide D. Ye}
\affiliation{Elmore Family School of Electrical and	Computer Engineering, Purdue University, West Lafayette, Indiana 47907, United States}
\affiliation{Birck Nanotechnology Center, Purdue University, West Lafayette, Indiana 47907, United
	States}
\author{J. Wunderlich}
\affiliation{Physics Department, University of Regensburg, 93040 Regensburg, Germany}

\author{S. D. Ganichev}
\affiliation{Physics Department, University of Regensburg, 93040 Regensburg, Germany}
\affiliation{CENTERA Labs, Institute of High Pressure Physics, PAS, 01 - 142 Warsaw, Poland}
\email{sergey.ganichev@ur.de}

		\begin{abstract}
We study terahertz field-driven second-order nonlinear electron transport phenomena, including the nonlinear Hall effect (NLHE), in two-dimensional tellurene flakes. The dc current excited by linearly polarized terahertz (THz) radiation in Hall bar samples is investigated in directions both along and perpendicular to the $c$-axis of tellurene. As expected for second-order transport phenomena, 
the current scales as the square of the in-plane electric field of the radiation $\bm E$, and depends on its orientation. 
The current results from a combination of three contributions, including the NLHE, the Nonlinear Longitudinal (NLL) and Nonlinear Diagonal (NLD) currents.
We established the equivalence between NLH, NLL, and NLD  transport currents and Linear photogalvanic effect (LPGE) contributions induced by the absorption of linearly polarized and unpolarized THz radiation. All contributions can be controlled by a gate voltage and have opposite signs for electron and hole conductivity. The magnitude of the current increases drastically when the samples are cooled from room temperature to 4.2~K. It also increases with decreasing radiation frequency. These results are well described by the developed phenomenological and microscopic theories. We show that the THz radiation-induced electric current originates from microscopic mechanisms such as skew scattering, side jump, and the Berry curvature dipole.
		\end{abstract}
	
	\maketitle	
	

	\section{Introduction}
	
Tellurene, the two-dimensional (2D) allotrope of trigonal tellurium, has recently emerged as an experimentally accessible elemental semiconductor, enriching the family of low-dimensional van der Waals materials with a structurally and functionally distinct system~\cite{Wu2018b,Wu2018a,Shi2020,Qiu2022,Zha2024}. Since trigonal Te is chiral and lacks inversion symmetry, tellurene hosts chirality-dependent electronic and optical responses and provides a natural platform for studying symmetry-allowed nonlinear transport. In $n$-type material, Weyl points are located near the band edge and together with the associated Berry-curvature hot spots make tellurene particularly attractive in the context of topological quantum materials~\cite{Qiu2020,Kim2024}. Beyond its topological aspects, tellurene combines high environmental stability and pronounced catalytic activity~\cite{Wu2017}, gate-tunable carrier density and carrier type~\cite{Wu2018b}, a strain-dependent band structure~\cite{Niu2023b}, significant piezoelectricity ~\cite{Apte2021,Rao2022}, a strongly anisotropic photoresponse~\cite{Gao2019}, low thermal conductivity, and room-temperature carrier mobilities 
close to
$10^3$~cm$^2$/Vs~\cite{Wang2018}. A variety of intriguing transport and optical phenomena have already been reported in tellurene, including the quantum Hall effect~\cite{Qiu2020,Niu2021}, spin Hall effect~\cite{Sachdeva2023}, weak antilocalization~\cite{Niu2020}, chirality-dependent nonlinear magnetotransport~\cite{Niu2023,Ma2024,SuarezRodriguez2025,Fontana2025,Neto2025,Iacovelli2026}, and THz helicity-sensitive currents~\cite{Moench2025}.

Most recently, the nonlinear Hall effect (NLHE) has been observed in transport measurements on tellurium-based thin flakes and 2D tellurene devices ~\cite{Cheng2024,Kim2024}. The NLHE is a transverse dc electric current or voltage response that is quadratic in the driving electric field, which appears even in time-reversal-symmetric conditions provided the material lacks inversion symmetry~\cite{Sodemann2015,Zhang2021a,Du2021,Du2021a,Ortix2021}. NLHE has been detected in topological insulators, Weyl semimetals, as well as 2D systems like transition metal dichalcogenides, graphene, topological surface states and others~\cite{Ma2019,Shen2025,Huang2022a,Huang2023,Jiang2025}. The sun-shape sample configuration has proven to be the most efficient method of detection~\cite{Kang2019}.  It is commonly emphasized that it is the Berry curvature dipole (BCD), which is responsible for microscopic origin of NLHE~\cite{Sodemann2015, Ma2019, Kang2019,Du2021, Du2021a, Zhang2021a,Ortix2021,Huang2022a, Huang2023, Luo2024, Cheng2024,Kim2024,Jiang2025,Shen2025}. At the same time, the BCD is not necessarily the dominant mechanism of NLHE, and other mechanisms, such as side-jump or asymmetric (skew) scattering, may play an essential role~\cite{Du2021,Du2021a,Ortix2021}. This distinction is especially relevant for tellurium-based systems: NLHE in gate-tuned 2D Te has been interpreted in terms of a large BCD near the Weyl band edge~\cite{Kim2024}, whereas room-temperature Te thin flakes have shown a giant NLHE dominated by extrinsic scattering and enabled by symmetry reduction at the surface~\cite{Cheng2024}.

Independent of the microscopic origin, the symmetry of the actual system must be stated carefully. In time-reversal-invariant systems, broken inversion symmetry is a necessary but not sufficient condition for the NLHE: the relevant second-order conductivity tensor must also contain symmetry-allowed components under the remaining crystal symmetries~\cite{Sodemann2015,Zhang2021a,Du2021,Du2021a,Ortix2021}. In substrate-supported 2D tellurene the symmetry is as low as $C_1$, and all in-plane second-order coefficients are symmetry-allowed. Consequently, transverse, longitudinal, and diagonal
in respect to the electric field
 second-order currents, as well as polarization-independent and helicity-dependent rectification, are not excluded by symmetry~\cite{Moench2025}.

\begin{figure*}[t]
	\centering
	\includegraphics[width=\linewidth]{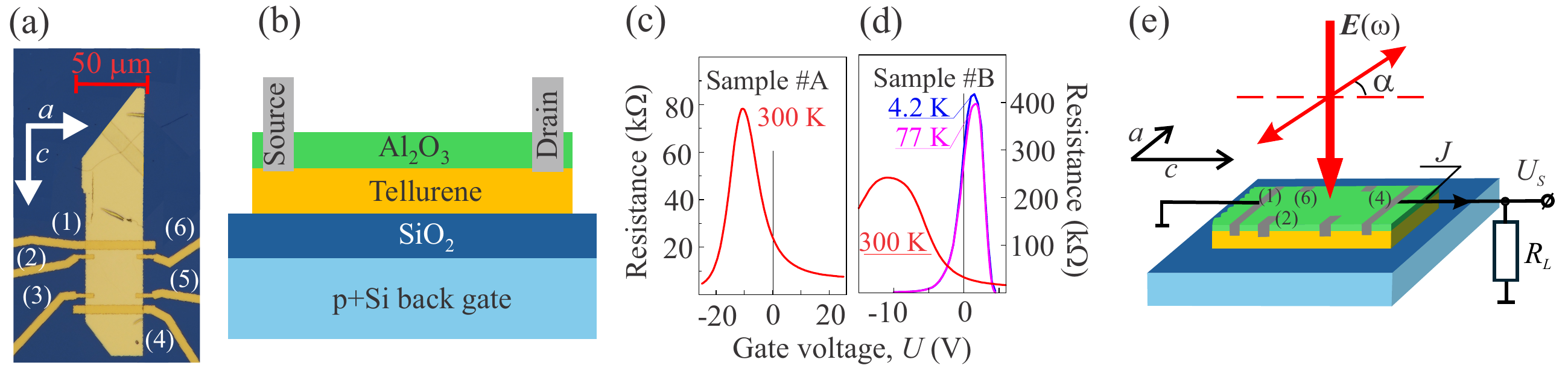}
	\caption{Panels (a) and (b) show  an optical micrograph of sample \# A and a cross-section of the samples. 
	The tellurene flake is deposited on top of $p+$ Si substrate covered by a 90~nm SiO$_2$ layer. Nickel  source and drain contacts with thickness of 50~nm  are fabricated by electron beam evaporation on the short sides of the tellurene flakes, which were $38 \times 41~\mu$m$^2$.  Te flakes are capped by a 15-nm-thick layer of Al$_2$O$_3$. Panels (c) and (d) show the two-point resistance measured along the $c$-axis as a function of the back gate voltage for samples \#A and \#B, respectively. Panel (e) sketches the experimental set-up. The azimuth angle $\alpha$ defines orientation of the radiation electric field vector $\bm{E}$ in respect to the $c$-axis.
	}
	\label{fig1}
\end{figure*}

It is important to highlight the connection between the NLHE and the photogalvanic effect  in radiation induced high-frequency experiments. The photogalvanic effect represents a generation of a dc electric current under absorption of light in unbiased systems~\cite{Belinicher1980,Ganichev2003b,Ivchenko2005,Ganichev2005}.	In optical language, the photo-current response to linearly polarized radiation is the linear photogalvanic effect (LPGE), where "linear" refers to the polarization state of the radiation rather than to the order of the response. Under an alternating electric field, the generated LPGE dc current is 
governed by just the same second-order conductivity tensor that underlies low-frequency nonlinear transport response. Consequently, the LPGE current component perpendicular to the exciting ac electric field is the high-frequency counterpart of the NLHE, while its component parallel to the field corresponds to nonlinear longitudinal transport.

In this work, we present the observation and a comprehensive study of THz radiation-driven second-order transport in 2D tellurene, which persists up to room temperature. By varying the relative orientation of the high-frequency electric field and the detected dc photocurrent, we resolve transverse, diagonal, and longitudinal second-order responses, including the NLHE. Our theory incorporates both intrinsic and extrinsic microscopic mechanisms of nonlinear current generation, namely the BCD, side-jump, and skew-scattering contributions. Our THz radiation approach offers several experimental advantages: (i) contactless application of the alternating electric field, (ii) straightforward control of the field orientation by a 
half-wave plate, (iii) direct access to dc longitudinal second-order currents without a linear dc background, (iv) implementation of  a simple Hall-bar geometry, and (v) operation in a large temperature range up to room temperature. 

\begin{figure}[t]
	\centering
	\includegraphics[width=\linewidth]{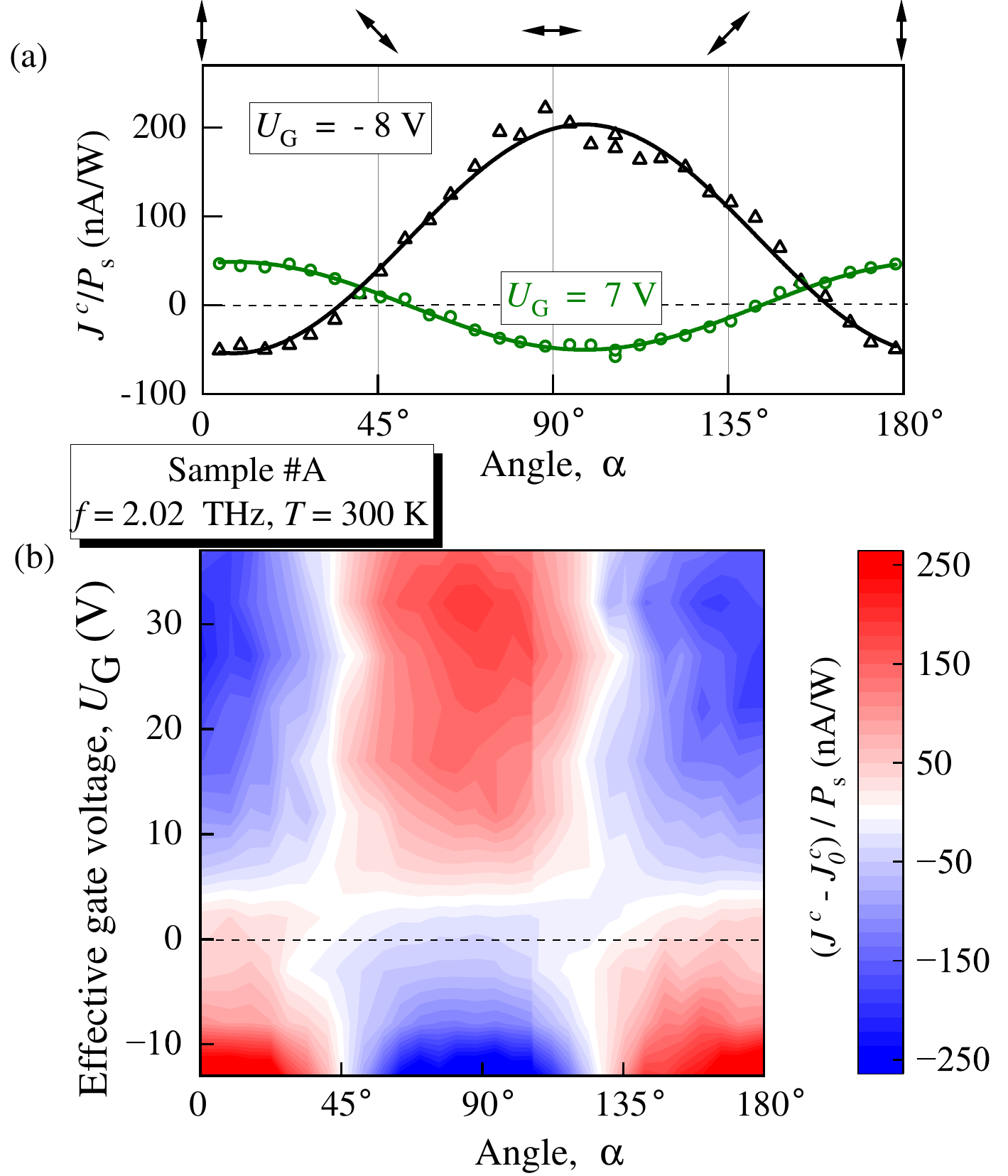}
	\caption{(a) Azimuth angle dependence of the photocurrent $J$ normalized on the power $P_s$ excited by radiation with $f=2.02$~THz. The data are shown for sample \#A at room temperature 
and two effective back gate voltages $U_{\rm G} = -8$~V and 7~V. Solid lines are fits after Eq.~\eqref{phenom1}. Double-arrows on top show orientations of the electric field vector for several angles $\alpha$.
	(b) Azimuth angle and effective gate voltage dependence of the photocurrent $(J^c-J_0^c)/P_s$ excited in sample \#A by radiation with $f=2.02$~THz, where $J_0$ is the polarization independent offset. The effective gate ranges negative (hole-conductivity) to positive values (electron-conductivity).
	}
	\label{fig2}
\end{figure}

\begin{figure}[t]
	\centering
	\includegraphics[width=\linewidth]{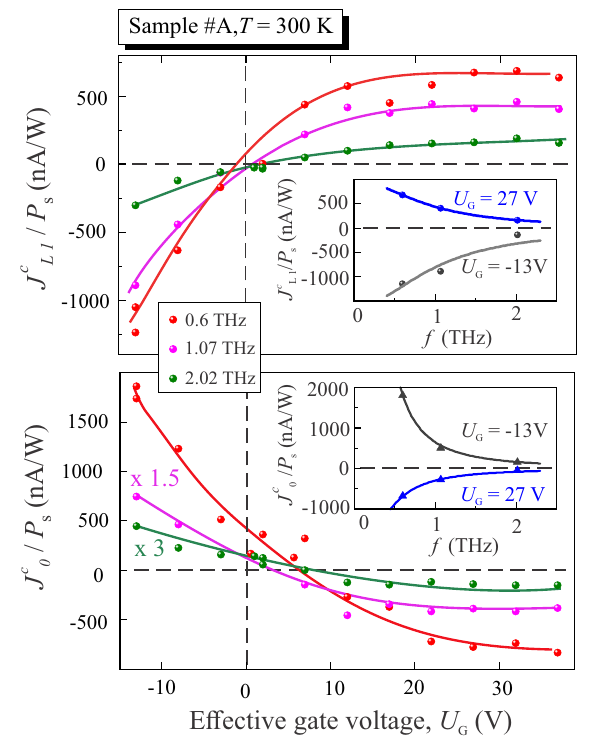}
	\caption{Effective gate voltage dependencies of the fitting parameters $J_{\rm L1}$, panel (a), and $J_0$, panel (b).  The data are obtained for sample \#A  excited by radiation of three frequencies.  The insets show  frequency  dependencies of $J_{\rm L1}$ and $J_0$ for several values of positive and negative effective gate voltages. Solid lines are guides for eye. 
		}
	\label{fig3}
\end{figure} 
\section{Phenomenological consideration}

We begin with the phenomenological theory of the nonlinear transport in 2D tellurene samples. Second order transport phenomena including the nonlinear Hall currents that are quadratic in the external electric field can appear even in time-reversal symmetric conditions. The direct current generated in response to a static electric field $\bm E$ is expressed by the following relation:  
\begin{align}
	\label{phen_dc}
	j^\alpha &= \sigma_{\alpha\mu}^{(1)}E_\mu + \sigma_{\alpha\mu\nu}^{(2)}E_\mu E_\nu ,
\end{align}
Here $\sigma_{\alpha\mu}^{(2)}$ and $\sigma_{\alpha\mu\nu}^{(2)}$ are the first (linear conductivity) and  second (nonlinear conductivity) order in electric field dc conductivity tensors, respectively, and $\alpha$, $\mu$, and $\nu$ run over in-plane Cartesian coordinates. In the high-frequency regime the first term in  Eq.~\eqref{phen_dc} averages to zero and the resulting dc current arises as a second-order response to the ac electric field $\bm E_\omega(t) = \bm E \exp(-i\omega t)+\bm E^* \exp(i\omega t)$. 
The dc current generated by linearly polarized radiation is sensitive to the orientation of the in-plane electric field and is given by
\begin{align}
	\label{phen_ac}
	j^\alpha &= \sigma_{\alpha\mu\nu}^{(2)}(\omega)E_\mu E^*_\nu  \nonumber\\
	&= \chi_{\alpha\mu\nu}(\omega)(E_\mu E_\nu^*+E_\mu^* E_\nu)\,, 
\end{align}
where $\hat{\bm \chi}(\omega)$ is the third rank tensor. The symmetry of studied 2D tellurene is reduced to the point group $C_1$~\cite{Moench2025}. In such systems, for excitation with homogeneous radiation at normal incidence, Eq.~\eqref{phen_ac} assumes the form 
\begin{align}
	\label{phenom1}
	J^c = J^c_{L1}  \cos{2 \alpha} + J^c_{L2} \sin{2 \alpha} + J^c_{0},
\\
J^a = J^a_{L1}  \cos{2 \alpha} + J^a_{L2} \sin{2 \alpha} + J^a_{0}.
	\label{phenom2}
\end{align}
Here $J^c \propto  \abs{\bm E}^2\propto P_s$ and $J^a\propto  \abs{\bm E}^2\propto P_s$ are currents excited along $c$- and $a$-axes ($a \perp c$), 
$J_{L1}^c$, $J_{L1}^a$ $J_{L2}^c$, $J_{L2}^a$, $J_{0}^c$ and $J_{0}^a$
 are six linearly-independent current contributions which, due to the absence of any nontrivial symmetry operation, are not related to one another; 
 $P_s $ is the radiation power, 
 and the azimuth angle $\alpha$ defines the orientation of the electric field in respect to the $c$-axis.

\section{Devices and experimental technique}

Experiments on THz nonlinear transport in two-dimensional (2D) tellurene samples  were performed on Hall bar samples with a long side oriented along the $c$-axis. A high-frequency electric field was applied to the sample via contactless 
normal-incident irradiation using a monochromatic THz laser,
with frequencies ranging from 0.6 to 2.54 THz. DC electric current caused by the nonlinear transport effects was measured along the $c$-axis (source-drain contacts) and $a$-axis (contacts 2 and 6 on the opposite sides of the Hall bar), see Figs.~\ref{fig1}(a) and~(e). The Te flakes were fabricated using a hydrothermal growth process similar to those previously reported for growing tellurene nanofilms~\cite{Qiu2020,Moench2025}.  The synthesized 2D Te flakes had a thickness of approximately 30~nm.  The as-grown 2D Te films were first cleaned twice with de-ionized water and then dispersed onto 
a 90~nm SiO$_2$/525~\textmu{m} p+Si  substrate using the Langmuir-Blodgett method~\cite{Zasadzinski1994}. 
Hall-bar devices were patterned using electron beam lithography, and 50~nm Ni metal contacts were deposited by electron beam evaporation at a pressure below $2\times 10^{-6}$~Torr. 
Figures~\ref{fig1}(a) and~(b) show the photograph and cross-section of the studied devices, respectively. The source and drain contact distance is about 40~$\mu$m. The substrate served as a back gate which allows to change the concentration and carrier type. We studied two samples (\#A and \#B) with close parameters. Sweeping the  gate voltage  $U$  shows  the charge neutrality point (CNP),  see Figs.~\ref{fig1}(c) and~(d). At room temperature the CNP is detected at large negative gate voltages, $U \approx -12$~V for the sample~\#A and $U \approx -10$~V for the sample~\#B. Thus for convenience we use here and below an effective gate voltage  $U_{\rm G} = U - U_{\rm CNP}$, where $U_{\rm CNP}$ is the voltage at which the CNP appears. For positive and negative $U_{\rm G}$ we have electron and hole-type conductivity, respectively~\cite{Niu2023a}. The carrier concentration for an effective gate voltage of $\pm 20$~V is about $5 \times 10^{12} $~cm$^{-2}$, and the room temperature mobility $\mu$ is about $430$ and 170~cm$^2$/Vs for holes and  electrons, respectively.

\begin{figure}[t]
	\centering
	\includegraphics[width=\linewidth]{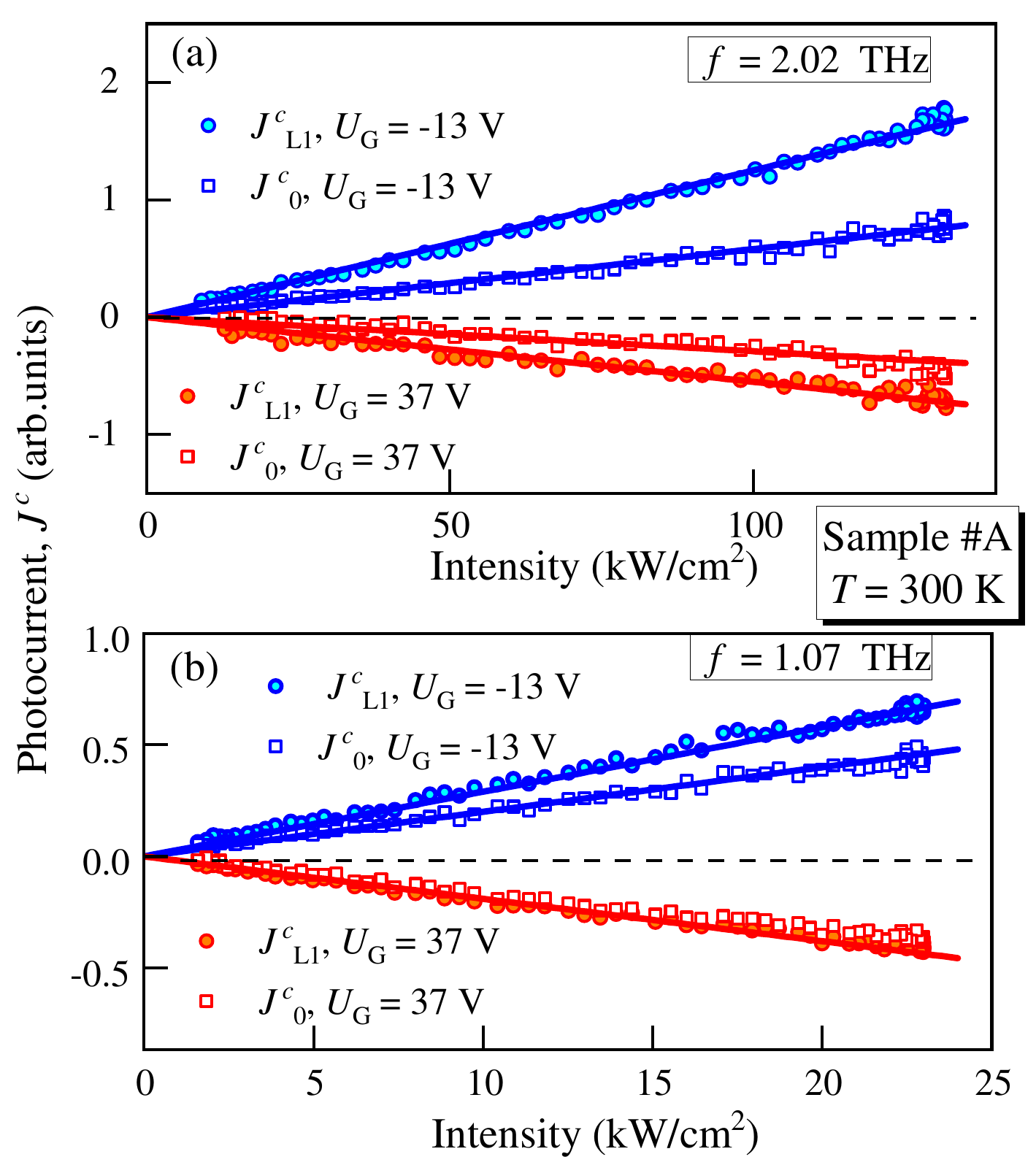}
	\caption{Dependencies of the photocurrent $J$ on the radiation intensity $I = P/S \propto E^2$. The data are obtained for frequencies $f = 2.02$ (upper panel) and $1.07$~THz (lower panel) and several effective gate voltages. Solid lines show fits after $J \propto E^2$. 
	}
	\label{fig4}
\end{figure} 

Pulsed and  continuous wave (cw) line-tunable  molecular THz lasers optically pumped by CO$_2$ lasers were used as a source of THz electric field. The pulsed NH$_3$ laser~\cite{Ziemann2000} pumped by a transversely excited atmosphere (TEA) CO$_2$ laser, operated at frequencies $f = 0.60$, $1.07$, and $2.02$\,THz  ($\hbar \omega = 2.5$, 4.4 and 8.35\,meV). The pulsed laser generated single pulses at a repetition rate of 1~Hz, with a pulse duration of approximately 100 ns. The laser's  peak power  and pulse shape were controlled by the THz photon drag detector~\cite{Ganichev1985}. Maximum peak power $P$ were about 20~kW (2.02~THz), 5~kW (1.07 THz) and 1.5~kW (0.6~THz). The cw CH$_3$OH laser operated at frequency 2.54~THz ($\hbar\omega=10.5$\,meV).  The radiation with the  power $P$ in the range from 5 to 30\,mW was modulated by an optical chopper operating at a frequency of about 80~Hz. The samples were exposed to linearly polarized radiation with normal incidence, see Fig.~\ref{fig1}(e), 
and the radiation induces indirect optical transitions (Drude-like free-carrier absorption) in the lowest subband.

Most experiments are performed at room temperature. To test the influence of temperature reduction, several experiments were also performed at 50~K and 4.2 K. For room temperature we used a 90~degree off-axis parabolic mirror with the focal plane of 75 mm. For low temperature measurements the samples were placed in a temperature controllable optical cryostat with $z$-cut quartz and TPX (4-methyl-1-pentene) windows. In this case the off-axis  mirror focal plane was 170~mm. In order to block visible and near-infrared radiation, the windows were additionally covered by black polyethylene foil. 
The beam shape and parameters were controlled using a pyroelectric camera~\cite{Ganichev1999,Ziemann2000}.  The pulsed laser spot diameters at FWHM varied from 2.6 to 3.1~mm depending on the radiation frequency. For cw laser radiation it was   0.9~mm. Considering the area of the devices, $S_{\rm device}$, we obtain that the power irradiating the sample $P_s = P\times (S_{\rm device}/S_{\rm spot})$, where $P$ is the radiation power, and $S_{\rm spot}$ is the  beam spot area. The orientation of the electric field with respect to the $c$-axis is defined by an azimuth angle,  $\alpha$, as shown in the inset of Fig.~\ref{fig1}(e). The angle $\alpha$  was controllably varied by rotating half-wave plates or by placing a grid wire behind a quarter-wave plate set. 
The radiation power dependencies were obtained using the cross-polarizer technique, which employs two wire grid polarizers. The first polarizer was rotated to modify the radiation power, while the second polarizer remained fixed to ensure an unchanged output polarization~\cite{Hubmann2019, Candussio2021a}.

In the set-up with pulsed laser the induced photocurrent was detected as a voltage drop $V_\text{ph}$ across load resistors  $R_\mathrm{L}=50$~Ohm. The signals were amplified by a voltage amplifier with 
gain of $k = 100$ and  a bandwidth of 300~MHz and recorded by digital  broad-band (1~GHz) oscilloscope. The photocurrent is obtained as $J = V_\text{ph} / (k R_\parallel)$, where $R_\parallel = (R_{\rm s} \times R_L) / (R_{\rm s} + R_L)$ and  $R_{\rm s}$ is the sample resistance. In the $cw$ laser setup the photovoltage signals $V_\text{ph}$ were measured by modulating the incoming radiation with an optical chopper and  using a standard lock-in technique.
The resulting signal is related to the photocurrent as $J = V_\text{ph} / ( k_{\rm a} R_\mathrm{s})$, where $k_{\rm a}$
is the preamplifier gain coefficient.

\begin{figure*}[t]
	\centering
	\includegraphics[width=\linewidth]{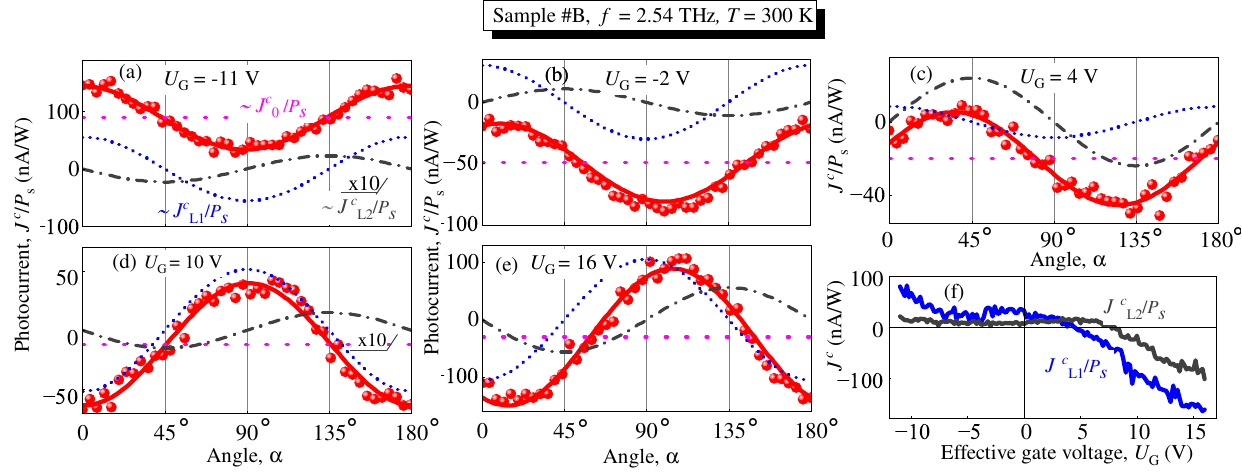}
	\caption{Panels (a)-(e): Dependencies of the normalized photocurrent $J^c/P_s$ (red circles) on the azimuth angle $\alpha$ measured in sample \#B. The data are  obtained for  $f = 2.54$~THz, room temperature, and for several effective back gate voltages ranging from hole ($U_{\rm G} < 0$) to electron conductivity ($U_G > 0$). The magnitudes of the currents  $J_{L2}^c$ in panels (a) and (d) are multiplied by factor 10 for visibility. Red solid lines are fits after Eq.~\eqref{phenom1}. The azimuth angle dependencies of the  individual current contributions $J_{L1}^c$, $J_{L2}^c$, and $J_{0}^c$ are plotted by blue dotted, dark gray dot-dashed and magenta dashed lines, respectively. Panel~(f): Current contributions $J_{L1}^c$, and $J_{L2}^c$ as a function of the effective gate voltage.
	}
	\label{figS1}
\end{figure*}

\begin{figure}[t]
	\centering
	\includegraphics[width=\linewidth]{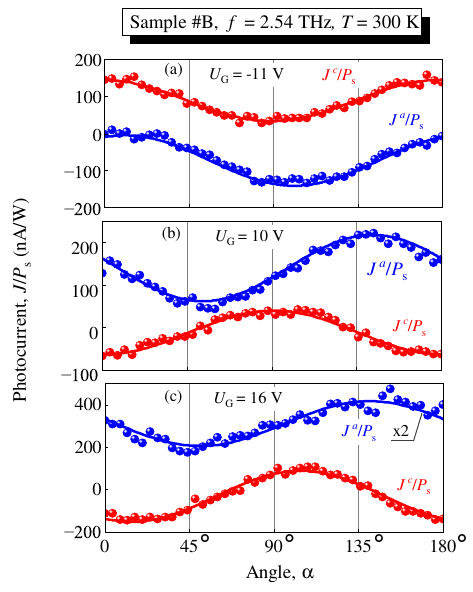}
	\caption{Dependencies of the normalized photocurrent $J^c/P_s$ (red circles) and $J^a/P_s$ (blue circles) on the azimuth angle $\alpha$ measured in sample \#B. The data are  obtained for  $f = 2.54$~THz, room temperature, and for several effective back gate voltages ranging from hole ($U_{\rm G} < 0$) to electron conductivity ($U_G > 0$).  The magnitudes of current $J^a/P_s$ in panel (c) are multiplied by a factor 2 for visibility. Red and blue solid lines are fits after Eqs.~\eqref{phenom1} and \eqref{phenom2}, respectively. 
	}
	\label{figS2}
\end{figure}

\section{Results}

First, we present the results obtained at room temperature when applying the pulsed laser. Figure~\ref{fig2}(a) shows the dependence of the dc current $J^c$ generated along the $c$-axis in sample \#A by the ac electric field with frequency $f=2.02$~THz. For positive effective gate voltages (circles) the current is excited in the conduction band and for the negative $U_{\rm G}$ in the valence band. The polarization dependence is well described by Eq.~\eqref{phenom1}. The fit reveals that the signal is dominated by the first ($\propto J^c_{\rm L1}$) and last terms ($\propto J^c_0$) of Eq.~\eqref{phenom1}. Figure~\ref{fig2}(b) shows a color map of the polarization dependent current contribution $J^c - J^c_0$ as a function of the azimuth angle $\alpha$ and effective gate voltage $U_{\rm G}$. The figure shows that, for all gate voltages, the polarization dependence of the current in sample \#A is determined by the first term of Eq.~\eqref{phenom1}. The same result has been obtained for lower frequencies 1.07 and 0.6~THz. Figures~\ref{fig3}(a) and (b) show the gate voltage dependencies of $J_{\rm L1}^c$ and $J^c_0$ contributions obtained for three radiation frequencies. These figures show that the magnitude of both contributions increase with the frequency decrease. This tendency is shown in the insets to Fig.~\ref{fig3} for two values of effective gate voltages  ($U_{\rm G}=-13$ and 27~V). Figure~\ref{fig4} shows the dependence of the total current $J^c$ on the electric field square obtained for frequencies $f=1.07$ and $2.02$~THz, 
and two gate voltages $U_{\rm G} = -13$ and 37~V. The data confirm that, in agreement with Eq.~\eqref{phenom1}, the current scales linearly with the radiation intensity $I = P/S_{\rm spot}\propto E^2$. 

While in the sample \#A at room temperature the current $J^c$ is dominated by the contributions proportional to the coefficients  $J_{\rm L1}^c$ and $J^c_0$ in Eq.~\eqref{phenom1} and the 
current defined by the coefficient $J_{\rm L2}^c$ is negligible, in sample \#B the latter one becomes essential for both positive and negative effective gate voltages.  Figures~\ref{figS1}(a)-(e) show the azimuth angle dependencies obtained for sample \#B excited by low power radiation of cw laser with frequency 2.54~THz. The data  are presented for several gate voltages. The data are well fitted by Eq.~\eqref{phenom1}, see solid lines in panels (a)-(e). Azimuth angle dependencies of the individual contributions proportional to the coefficients $J_{\rm L1}^c$, $J_{\rm L2}^c$ and $J^c_0$, which were used for the fits, are shown by dashed, dot-dashed and dotted lines, respectively. It is seen that for some effective gate voltages the contribution $\propto J_{\rm L2}^c$ is substantial and can even dominate, see, e.g., Fig~\ref{figS1}(c). 

The current is also detected in the direction $a$ perpendicular to the $c$-axis, see Figs.~\ref{figS2}(a)-(c). It is seen that the current magnitudes for these two directions are very close to each other. We also note that  for this direction the 
contribution $\propto J_{\rm L2}^c$ varying as $\sin(2\alpha)$ is  dominating for positive effective gate voltages.  

\begin{figure*}[t]
	\centering
	\includegraphics[width=\linewidth]{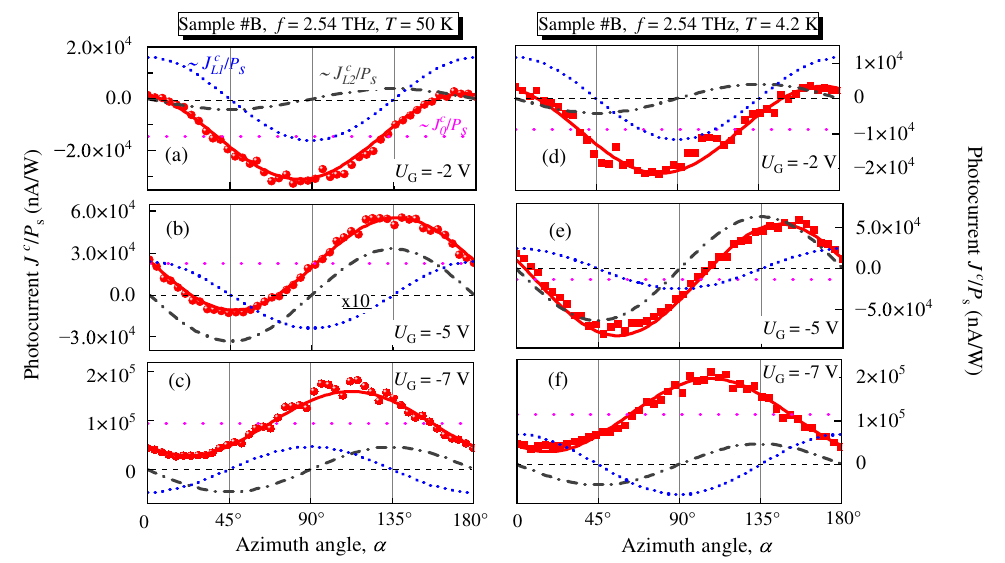}
	\caption{Dependencies of the normalized photocurrent $J^c/P_s$ (circles) on the azimuth angle $\alpha$ measured in sample \#B for source-drain pair of contacts and several effective back gate voltages. The data are obtained for  $f = 2.54$~THz and two temperatures 50~K, panels (a)-(c), and 4.2~K, panels (d)-(f). The magnitudes of the currents  $J_{L2}^c$ in panel (b) are multiplied by factor 10 for visibility. 		Solid lines are fits after Eq.~\eqref{phenom1}. The azimuth angle dependencies of the  individual current contributions $J_{L1}^c$, $J_{L2}^c$, and $J_{0}^c$ are plotted by blue dotted, dark gray dot-dashed and magenta dashed lines, respectively.
	}
	\label{fig9}
\end{figure*}

All data presented so far have been obtained in samples at room temperature. Cooling down the tellurene devices results in substantial (by two orders of magnitude) increase of the current magnitude. Figures~\ref{fig9}(a)-(c) and (d)-(f) show the data obtained in sample \#B for temperatures 50 and 4.2~K and radiation frequency 2.54~THz (low power cw-laser). Alike at room temperature, see Fig.~\ref{figS1}, the data can be well fitted by Eq.~\eqref{phenom1} with comparable magnitude of all contributions.

\section{Microscopic theory}

\label{Theory}

As shown above, the functional dependencies of the current excited by THz radiation can be well described by the phenomenological theory of second-order transport phenomena. In agreement with Eqs.~\eqref{phenom1} and~\eqref{phenom2}, the current scales as the square of the radiation electric field and varies with the azimuth angle as a sum of $\cos(2\alpha)$ and  $\sin(2\alpha)$, as well as has a polarization-independent contribution, see Figs.~\ref{fig2},  \ref{fig4}, \ref{figS1},  and \ref{fig9}. This behavior is observed in directions both along and perpendicular to the  $c$-axis, see Fig.~\ref{figS1}. Together with the fact that the current is detected at normal incidence, these observations confirm the reduction of symmetry compared to bulk Te, and indicate that the symmetry of the studied 2D tellurene is  $C_1$, ruling out any nontrivial symmetry elements. Analyzing the current behavior in relation to variations in microscopic parameters such as gate voltage and radiation frequency
requires a microscopic theory of nonlinear transport, which is provided below.   

The investigated LPGE currents are excited by THz radiation with a photon energy of a few meV, which is lower than the energy gap of tellurene. The typical carrier energies also exceed the photon energy. Consequently, LPGE is caused by the free carrier Drude-like radiation absorption in the conduction or valence band. 
Therefore, a theoretical description of the photocurrents is performed in the semiclassical picture in which  the THz radiation acts on the carriers as an ac electric field~\cite{Olbrich2014,Otteneder2020,Moench2025}
\begin{equation}
\bm E_\omega(t) = \bm E \exp(-i\omega t)+c.c.
\end{equation}

In the semiclassical theory, transport is described by the Boltzmann kinetic equation for the electron distribution function $f_{\bm k}$. Here $\bm k$ is the two-dimensional electron wavevector, and the equation has the following form
\begin{equation}
	\label{kin_eq}
	{\partial f_{\bm k}\over \partial t} + {q \over \hbar}\bm E_\omega(t) \cdot {\partial f_{\bm k}\over \partial \bm k} = \sum_{\bm k'} (W_{\bm k \bm k'}f_{\bm k'}-W_{\bm k' \bm k}f_{\bm k}),
\end{equation}
where $W_{\bm k' \bm k}$ is the probability of $\bm k' \leftarrow \bm k$ electron scattering,
and $q$ is the carrier charge
being $\pm \abs{e}$ for holes and electrons, respectively.
The electric current density $\bm j$ is calculated as 
\begin{equation}
\label{current_density}
\bm j = g_s g_v q\sum_{\bm k}f_{\bm k} \bm v_{\bm k},
\end{equation} 
where $g_s$ and $g_v=2$ are the spin and valley degeneracies,
and $\bm v_{\bm k}$ is the carrier velocity. 
For electrons and holes in tellurene we have $g_s=2$ and $g_s=1$, respectively.

There are three microscopic mechanisms of THz radiation-induced photocurrents. These are the extrinsic contributions due to skew scattering processes and side jumps, as well as the intrinsic Berry curvature dipole mechanism~\cite{Deyo2009,Golub2020,Du2021,Du2021a,Ortix2021}.
These mechanisms completely exhaust the photocurrent driven by circular polarization, as proven by a comparison with a rigorous quantum-mechanical calculation results~\cite{Golub2020} -- at least in gapped systems. However, the gapless Weyl semimetals require further investigation~\cite{Golub2026a}.

\begin{figure}
	\centering
	\includegraphics[width=\linewidth]{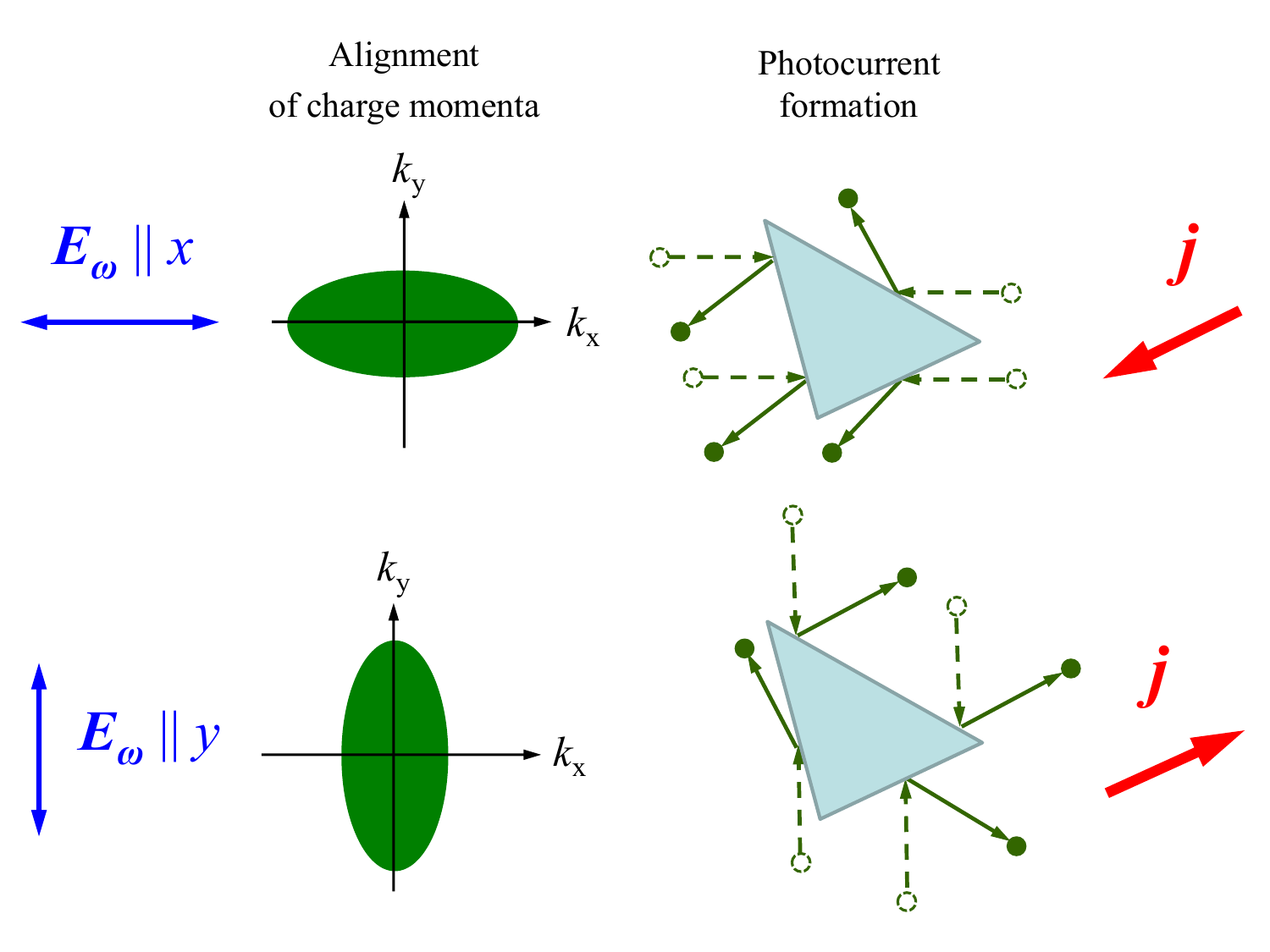}
	\caption{Microscopic mechanism of the photocurrent formation. Under linearly polarized THz excitation, the carrier momentum distribution becomes anisotropic: the number of particles with momenta aligned parallel or antiparallel to the electric field exceeds that of carriers with transverse momenta. The corresponding stationary momentum distribution is shown in green. Scattering of these momentum-aligned carriers (dashed arrows) by irregular triangles results in uncompensated charge flows (solid arrows) and in a dc electric current (red arrows). The photocurrent is shown for holes,
	it reverses its direction for electrons.
	}
	\label{Fig_triangles}
\end{figure}

First we consider the skew-scattering contribution.
The generation process of the skew-scattering induced LPGE can be illustrated as follows. At linear polarization of THz radiation, an alignment of electron momenta takes place: there is a larger number of particles with momenta along and opposite to the electric field direction and a smaller number in the perpendicular direction. The corresponding stationary momentum distribution is shown in Fig.~\ref{Fig_triangles} (left panels) by green color. Due to low ($C_1$) symmetry of the tellurene sample, scattering of 2D carriers occurs by asymmetrical defects. Scattering of the momentum-aligned carriers (dashed arrows) by such defects results in uncompensated charge flows (solid arrows) and in a dc electric current (red arrows). The photocurrent is shown for holes,
it reverses its direction for electrons.

Microscopically, the photocurrent, being a second-order dc response, is given
by the correction to the distribution function obtained by two iterations of the kinetic Eq.~\eqref{kin_eq} in powers of the electric field $\bm E$. In this approach, the distribution function is sought in the form 
\begin{equation}
f_{\bm k}=f_0+[f_1\exp(-i\omega t) + c.c.] + f_2,
\end{equation}
where $f_0$ is the equilibrium distribution at $E=0$, $f_1 \propto E$ is a linear correction, and $f_2 \propto E^2$ is a dc contribution quadratic in $E$.
Double iteration of the Boltzmann kinetic Eq.~\eqref{kin_eq} in the electric field amplitude $\bm E_\omega(t)$ yields the stationary correction to the charge carrier's distribution function $f_2(\bm k)$ describing an alignment of their momenta. The alignment means a larger number of carriers with momenta along $\pm \bm E_\omega$ direction than perpendicular to it. The dc part of the distribution function, $f_0+f_2$, is shown by green color in Fig.~\ref{Fig_triangles}. 

Momentum alignment does not describe an electric current because the corresponding distribution is even in $\bm k$. 
In order to obtain an odd in momentum part of the distribution, 
we take into account the skew scattering.

In low symmetry systems, the scattering probability in the kinetic Eq.~\eqref{kin_eq} does not satisfy the principle of detailed balance: the probability of $\bm k \leftarrow \bm k'$ process, $W_{\bm k \bm k'}$ differs from that of inverse process, $W_{\bm k' \bm k}$. Therefore it is convenient to make a splitting to the symmetric and asymmetric parts: $W_{\bm k \bm k'}= W_{\bm k \bm k'}^{s} + W_{\bm k \bm k'}^{a}$, where $W_{\bm k \bm k'}^{s,a} = \pm W_{\bm k' \bm k}^{s,a}$. The asymmetric part  $W_{\bm k \bm k'}^{a}$ is nonzero in systems of $C_1$ symmetry~\cite{Otteneder2020} and is responsible for the skew scattering.

The symmetrical part of the scattering probability $W_{\bm k \bm k'}^{s}$ defines the relaxation times of different Fourier harmonics of the distribution function. In particular, we introduce the times $\tau_{1,2}$ dependent on the carrier's energy $\varepsilon_k$
defined as follows~\cite{Olbrich2014}
\begin{equation}
\tau_{n}^{-1}(\varepsilon_k)=\sum_{\bm k'}W_{\bm k \bm k'}^{s}(1-\cos{n\theta_{\bm k \bm k'}}),
\end{equation}
where $\theta_{\bm k \bm k'}$ is the scattering angle.
At $\varepsilon_k$ equal to the Fermi energy $\varepsilon_{\rm F}$, they give the transport scattering time which determines the mobility, $\tau_{\rm tr}=\tau_1(\varepsilon_{\rm F})$ and the alignment relaxation time $\tau_{\rm al}=\tau_2(\varepsilon_{\rm F})$.

The momentum-aligned carriers are scattered by asymmetrical defects predominantly in the same direction, Fig.~\ref{Fig_triangles}, which means a net electric current. The current is microscopically calculated by taking into account $W_{\bm k \bm k'}^{a}$ in the lowest order. The corresponding dc correction to the distribution function, ${f}_2^{a}$, yields the current density by Eq.~\eqref{current_density} with the band velocity 
\begin{equation}
\label{v_band}
\bm v_{\bm k}^{\rm band}={1\over \hbar}\bm \nabla_{\bm k}\varepsilon_k.
\end{equation}
The LPGE current density driven by linear polarization of radiation reads
\begin{align}
	\label{j_micro}
&j_{c,a} = \abs{\bm E}^2 ev_\text{F}\sigma(\omega)\biggl\{
2{\left[v_\text{F}^3 \tau_{\rm al} (\Xi_{c,s}P_{\text{L1}}+\Lambda_{c,s}P_{\text{L2}})\right]'\over v_\text{F}^3} - {\tau_{\rm al}\over \tau_{\rm tr}}\nonumber\\
&\times \tau_{\rm tr}' {1-\omega^2\tau_{\rm tr}\tau_{\rm al}\over 1+\omega^2\tau_{\rm al}^2}
[(\Xi_{c,s}\mp\Lambda_{s,c})P_{\text{L1}} \mp (\Xi_{s,c}\pm\Lambda_{c,s})P_{\text{L2}}]
\biggr\}. 
\end{align}
Here $\sigma(\omega)=\sigma_0/(1+\omega^2\tau_{\rm tr}^2)$ is the frequency-dependent conductivity with $\sigma_0$ being the dc conductivity,
\begin{equation}
\label{Stokes_param}
P_{\text{L1}}={\abs{E_{c}}^2-\abs{E_{a}}^2 \over \abs{\bm E}^2}, \quad P_{\text{L2}}={2\text{Re}(E_{c}E_{a}^*)\over \abs{\bm E}^2}
\end{equation}
are the Stokes parameters describing the linear polarization degree of radiation, the prime denotes the derivative with respect to the Fermi energy, and $v_\text{F}$ is the Fermi velocity. We assumed a parabolic energy dispersion and degenerate statistics.
The asymmetry of the scattering probability present in the $C_1$ point group is taken into account by the dimensionless factors $\Xi_{c,s}, \Lambda_{c,s} \ll 1$ defined as follows
\begin{align}
	\label{Xi_Lambda}
	\Xi_{c(s)} = \tau_\text{tr} \sum_{\bm k'}\left< \cos{(2\varphi_{\bm k})} W^{a}_{\bm k' \bm k} \cos(\sin){\varphi_{\bm k'}}\right>, 
	\\
	\Lambda_{c(s)} = \tau_\text{tr} \sum_{\bm k'}\left< \sin{(2\varphi_{\bm k})} W^{a}_{\bm k' \bm k} \cos(\sin){\varphi_{\bm k'}}\right>, \nonumber
\end{align}
where the brackets denote averaging over the directions of $\bm k$ at the Fermi circle, and the polar angles are measured from the $c$-axis.

There are two terms in the skew scattering probability resulting in the LPGE current contributions depending differently on the density of scatterers. A so-called conventional skew scattering occurs with non-Gaussian disorder and appears in the third order in the scattering potential. It is present at low temperatures for scattering by impurities, but is absent for acoustic phonon scattering when single-phonon processes are dominant~\cite{Glazov2020,Glazov2020_PRL}. 
The corresponding contribution to the skew scattering probability contains the density of defects $\mathcal N_d$, therefore $\Xi_{c,s}$ and $\Lambda_{c,s}$ are independent of $\mathcal N_d$, and the LPGE current $j \propto 1/\tau \propto \mathcal N_d$ at $\omega\tau \gg 1$. Another, the so-called intrinsic (or coherent) skew scattering, occurs at any potential~\cite{Sinitsyn2007,Ado2015,Ado2017}. The intrinsic skew scattering probability is proportional to the fourth power of the disorder potential and to $\mathcal N_d^2$. It gives a contribution to LPGE which is scales quadratically with the scattering rates and $\mathcal N_d$ at high frequencies: $j \propto \mathcal N_d/\tau \propto \mathcal N_d^2$. For scattering by phonons, $\mathcal N_d$ is substituted by the temperature~\cite{Glazov2020}.

We note that, in contrast to the intraband photocurrent driven by circular polarization, where the skew scattering gives a small contribution at ${\omega\tau \gg 1}$~\cite{Golub2020}, the LPGE current can be effectively generated due to this extrinsic mechanism at all frequencies, and can be even dominant.

There is another mechanism of LPGE which is related with the Berry curvature of carriers in the conduction and valence bands of tellurene. 
In the semiclassical approach, the Berry curvature dipole (BCD) mechanism of LPGE is accounted for in the anomalous velocity linear in the radiation electric field and oscillating in time with the frequency $\omega$:
\begin{equation}
\bm v^{\rm anom}_{\bm k}(t)=\bm \Omega_{\bm k} \times \bm E_\omega(t).
\end{equation}
Here the Berry curvature reads $\bm \Omega_{\bm k} = \bm \nabla_{\bm k}\times i\braket{u_{\bm k}}{\nabla_{\bm k} u_{\bm k}}$, with $u_{\bm k}$ being the electron Bloch amplitude  for the energy band where the radiation is absorbed.
Finding the linear correction to the distribution function $f_1 \propto E_\omega(t)$ from the Boltzmann Eq.~\eqref{kin_eq} with the symmetrical part of the scattering probability $W^s_{\bm k' \bm k}$ only, the BCD contribution to the LPGE current is then calculated by Eq.~\eqref{current_density} with $\bm v_{\bm k}=\bm v^{\rm anom}_{\bm k}(t)$.
The corresponding dc current density reads~\cite{Deyo2009}
\begin{equation}
\label{BCD}
\bm j^{\rm BCD}=-{2q^3\tau\over 1+\omega^2\tau^2}[\hat{\bm z} \times \bm E] (\bm D \cdot \bm E).
\end{equation}
Here $\bm D$ is the Berry curvature dipole introduced in Ref.~\cite{Sodemann2015}:
\begin{equation}
\bm D = {2 g_s g_v \over \hbar^2}{\rm Im}\sum_{\bm k}f_0(\varepsilon_k) \bm \nabla_{\bm k} \braket{\partial_{k_x}u_{\bm k}}{\partial_{k_y}u_{\bm k}}.
\end{equation}

Finally, there are contributions to LPGE caused by side jumps of carrier's wavepackets occurring at momentum scattering~\cite{Deyo2009}. The side-jump $\bm r_{\bm k' \bm k}$ for the scattering process $\bm k' \leftarrow \bm k$ is given by
\begin{equation}
\bm r_{\bm k' \bm k} = -(\bm \nabla_{\bm k'}+\bm \nabla_{\bm k})\Phi_{\bm k' \bm k} + \bm A_{\bm k'} - \bm A_{\bm k},
\end{equation}
where $\Phi_{\bm k' \bm k}$ is the phase of the matrix element of scattering $U_{\bm k'\bm k}$, and the Berry connection $\bm A_{\bm k}= i\braket{u_{\bm k}}{\bm \nabla_{\bm k} u_{\bm k}}$.

The first contribution 
comes from the side-jump accumulation resulting in the scattering-induced correction to the velocity:
\begin{equation}
\bm v^{\rm sj}_{\bm k}=\sum_{\bm k'}W^s_{\bm k' \bm k}\bm r_{\bm k' \bm k}.
\end{equation}
Then finding the correction to the distribution function $f_2 \propto E^2$ from the Boltzmann Eq.~\eqref{kin_eq} with 
the scattering probability $W^s_{\bm k' \bm k}$, the side-jump accumulation contribution to the current is calculated by Eq.~\eqref{current_density} with $\bm v_{\bm k}=\bm v^{\rm sj}_{\bm k}$.
This contribution to the LPGE current is comparable to the BCD one. Note that the side-jump accumulation is insensitive to the radiation helicity and, hence, does not contribute to the circular polarization-driven current.

The second correction to the distribution function caused by side jumps comes from 
the linear in the electric field scattering probability $W^{\rm sj}_{\bm k' \bm k}(t)$:
\begin{equation}
W^{\rm sj}_{\bm k' \bm k}={2\pi\over\hbar} \mathcal N_d \abs{U_{\bm k'\bm k}}^2(e\bm E_\omega \cdot \bm r_{\bm k' \bm k}) \partial_{\varepsilon_{\bm k}}\delta(\varepsilon_{k}-\varepsilon_{k'}).
\end{equation}
It can be interpreted as a correction to the energy conservation law $\delta(\varepsilon_{k}+e\bm E_\omega \cdot \bm r_{\bm k' \bm k}-\varepsilon_{k'})-\delta(\varepsilon_{k}-\varepsilon_{k'})$
due to a work of the field at the side-jump~\cite{Sinitsyn2007a}.
Iterating the kinetic Eq.~\eqref{kin_eq} with $W_{\bm k' \bm k}=W^s_{\bm k' \bm k}+W^{\rm sj}_{\bm k' \bm k}(t)$ in $\bm E_\omega (t)$, one finds the so-called anomalous distribution $f^{\rm adist} \propto E^2$, which is time-independent and already odd in $\bm k$. Then the corresponding contribution to the LPGE current is calculated by Eq.~\eqref{current_density} with $f_{\bm k}=f^{\rm adist}$ and the band velocity $\bm v_{\bm k}^{\rm band}$ given by Eq.~\eqref{v_band}.

Comparing the conventional skew-scattering, coherent skew-scattering, BCD and side-jump contributions to the LPGE current, we see that the first one given by Eq.~\eqref{j_micro} is different from the others which have the same order of magnitude. The ratio of the currents~\eqref{j_micro} and~\eqref{BCD} can be estimated as~\cite{Glazov2020,Glazov2020_PRL}
\begin{equation}
\label{compar}
{j^{\rm skew} \over j^{\rm BCD}} \sim {\varepsilon_{\rm F}\over \mathcal N_d \abs{U_{\bm k'\bm k}} }.
\end{equation}
While it is hard to model disorder in the studied tellurene samples, we note that the Fermi energy at highest gate voltages applied is $\varepsilon_{\rm F} \approx 50$~meV. This allows us to assume that the conventional skew-scattering contribution Eq.~\eqref{j_micro} dominates in the LPGE current.

The high-frequency asymptotics $j \propto 1/\omega^2$ is a feature common for all microscopic mechanisms, see Eqs.~\eqref{j_micro} and~\eqref{BCD}. Analysis of Eqs.~\eqref{j_micro} and~\eqref{Xi_Lambda} shows that the LPGE current is an increasing function of the Fermi energy. Therefore, the current amplitude should increase with increasing electron and hole densities, i.e., with the absolute value of the gate voltage.

\section{Discussion}

The results of the microscopic theory describe the LPGE current driven by linearly polarized radiation, see Eq.~\eqref{j_micro}. Since the Stokes parameters~\eqref{Stokes_param} are varied in the experiments as 
\begin{equation}
P_{\text{L1}}=\cos{2\alpha}, \quad P_{\text{L2}}=\sin{2\alpha},
\end{equation}
the developed theory is in agreement with the phenomenological Eqs.~\eqref{phenom1} and~\eqref{phenom2}, as well as experimental results, see Figs.~\ref{fig2},  \ref{figS1}, \ref{figS2},  and \ref{fig9}.
In the system of $C_1$ point symmetry under study, all six current contributions $J_{\rm 0, L1, L2}^{c,a}$ are nonzero and linearly independent.
Indeed, a strong polarization dependence is clearly seen in Figs.~\ref{fig2},  \ref{figS1}, \ref{figS2},  and \ref{fig9}.
Importantly, our results reveal that the polarization-independent current components $J_0^{c,a}$ are comparable or even larger than $J_{\rm L1}^{c,a}$, see Figs.~\ref{fig2}, \ref{fig3}, \ref{figS1}, \ref{figS2} and \ref{fig9}.
Note that 
polarization-independent currents
$J^c_{0}$ and $J^a_{0}$ may differ 
due to the in-plane anisotropy of mobility.
Moreover, the  terms in Eqs.~\eqref{phenom1},~\eqref{phenom2} containing $J_{\rm L2}^{c,a}\sin{(2\alpha)}$, which have a maximum amplitude $J_{\rm L2}^{c,a}$ for the electric field $\bm E$ oriented at $45^\circ$ to the $c$ and $a$ axes, 
under some conditions are comparable to $J_{\rm L2}^{c,a}$ and $J_0^{c,a}$, see dotted-dashed lines in Figs.~\ref{figS1}, and \ref{fig9} as well as the data in Figs.~\ref{figS2}(b) and (c).

Microscopically, the skew-scattering, Berry curvature dipole and side-jumps contribute to all three LPGE currents $\bm J_{\rm 0,L1,L2}$. In particular, the BCD contribution, Eq.~\eqref{BCD} in components reads
\begin{equation}
\begin{bmatrix}
j^{\rm BCD}_c\\j^{\rm BCD}_a
\end{bmatrix}
=
{q^3\tau \abs{\bm E}^2/2\over 1+\omega^2\tau^2}
\begin{bmatrix}
D_c \sin{2\alpha} + D_a(1-\cos{2\alpha})\\ 
-D_c(1+\cos{2\alpha}) - D_a\sin{2\alpha}
\end{bmatrix},
\end{equation}
i.e. contains both the polarization-independent and two polarization-dependent contributions.
The current $J_0^{c,a}$, in addition to the skew scattering, BCD and side jump mechanisms, may come from the electron gas heating by the electric field. 

\begin{figure*}
	\centering
	\includegraphics[width=\linewidth]{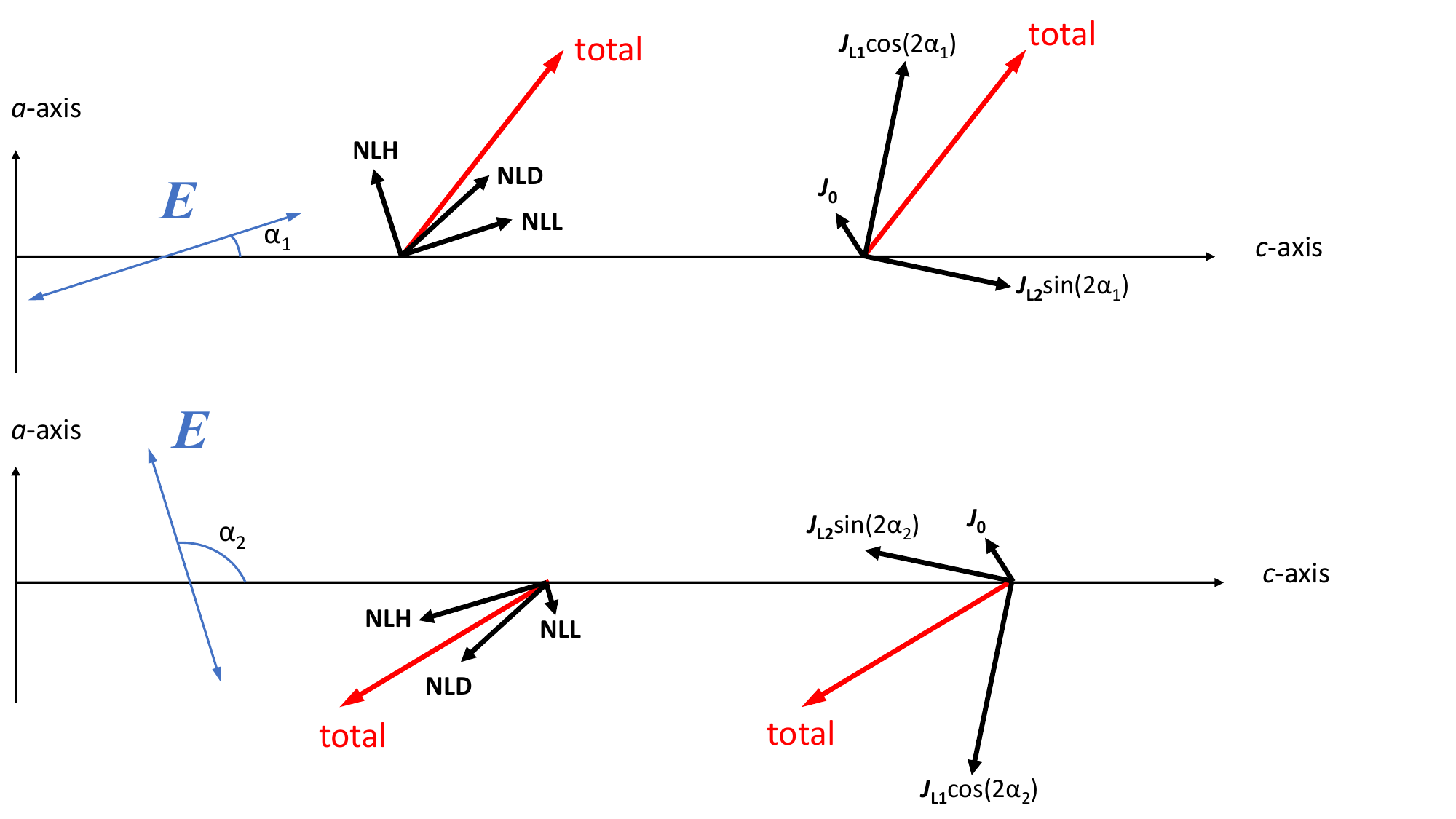}
	\caption{Different contributions to the LPGE current for two polarization directions. Three currents correspond to the NLH, NLL and NLD contributions in the transport representation and to the polarization-independent $\bm J_0$ and two polarization-dependent $\bm J_{\rm L1,L2}$ terms in the photogalvanic representation.
	}
	\label{fig_contributions}
\end{figure*}

In the  representation widely used for analysis of the second-order  transport phenomena, the LPGE current given by ~\eqref{phenom1},~\eqref{phenom2} is a sum of vectors perpendicular and parallel to $\bm E$ and -- in low symmetric systems -- of two additional terms
\begin{multline}
\label{J_transp_represent}
\bm J =
\begin{bmatrix}
J^c\\J^a
\end{bmatrix}
= 
\begin{bmatrix}
J_0^c + J_{\rm L1}^c \cos{2\alpha}+J_{\rm L2}^c \sin{2\alpha}
\\
J_0^a + J_{\rm L1}^a\cos{2\alpha}+J_{\rm L2}^a\sin{2\alpha}
\end{bmatrix}
\\ =[\hat{\bm z} \times \bm E] \qty(\bm C_{\rm NLH} \cdot \bm E)
+
\bm E \qty(\bm C_{\rm NLL} \cdot \bm E)
\\
+ C_{\rm NLD}^{(1)}  \bm S + C_{\rm NLD}^{(2)}  [\bm S\times \hat{\bm z} ],
\end{multline}
where $\bm S = (2E_cE_a,E_c^2 - E_a^2).$
The  term containing $[\hat{\bm z} \times \bm E]$ 
represents the NLHE current~\cite{Sodemann2015,Zhang2021a,Du2021,Du2021a,Ortix2021},  the contribution $\propto \bm E$ is the Nonlinear Longitudinal current (NLL)~\cite{Moench2025},
and two last terms represent the Nonlinear Diagonal (NLD) current discussed below.

The LPGE currents $J_{\rm 0, L1, L2}^{c,a}$ are related to the NLL and NLH coefficients $C_{\rm NLL,NLH}^{c,a}$ and new coefficients $C_{\rm NLD}^{(1,2)}$ by
\begin{align}
&C_{\rm NLL}^{c,a}\abs{\bm E}^2 =J_{\rm 0}^{c,a} + { J_{\rm L2}^{a,c}\pm  J_{\rm L1}^{c,a}\over 2}, 
\\
 &C_{\rm NLH}^{c,a}\abs{\bm E}^2= \pm J_{\rm 0}^{a,c} + { J_{\rm L1}^{a,c} \mp  J_{\rm L2}^{c,a}\over 2},
\\
&C_{\rm NLD}^{(1,2)}\abs{\bm E}^2={ J_{\rm L1}^{a,c}\pm J_{\rm L2}^{c,a}\over 2}.
\end{align}
The NLL, NLH and NLD contributions to the total current for two polarization directions are shown in Fig.~\ref{fig_contributions} together with a decomposition to the polarization-independent $\bm J_0$ and polarization-dependent $\bm J_{\rm L1,L2}$ LPGE currents.

The microscopics of NLHE is usually associated with the Berry curvature dipole (BCD) mechanism~\cite{Sodemann2015}. 
Indeed, the BCD contributes to $\bm j \perp \bm E$ only, see Eq.~\eqref{BCD}. 
At the same time, an essential contribution comes from the skew scattering~\cite{Deyo2009}. The corresponding current density is discussed above and given by Eq.~\eqref{j_micro}. Furthermore, the side-jump mechanism also gives rise to the NLHE~\cite{Deyo2009}.
The BCD and side jump mechanisms result in the comparable contributions to NLHE current whereas the skew scattering differs from them parametrically and can be dominant, see Eq.~\eqref{compar} and corresponding discussion. 
 Finally, we note that the electron gas heating also contributes to NLHE as it is defined in the literature, namely $\bm j \propto [\hat{\bm z}\times \bm E]$, see, e.g., Refs.~\cite{Sodemann2015,Zhang2021a,Du2021,Du2021a,Ortix2021}.
 
Furthermore, the theory, in agreement with the experiment, shows that the electric current is generated along the electric field direction $\bm E$, see the current values  in Figs.~\ref{fig2},  \ref{figS1}, \ref{figS2} and \ref{fig9} for angles 
 $\alpha=0$ ($J^c$) and $\alpha=90^\circ$ ($J^a$). 
 These contributions are partially caused by the Nonlinear Longitudinal current (NLL).
Microscopically, this current is caused by the skew scattering and side jump mechanisms, see
Sec.~\ref{Theory}.
Furthermore, alike the NLHE discussed above, $J_0^c$ contributes to the NLL current, which  may have a contribution caused   by the electron gas heating.

The coefficients $C_{\rm NLD}^{(1,2)}$ in Eq.~\eqref{J_transp_represent} describe the new effect in the nonlinear transport -- the Nonlinear Diagonal current. 
Accounting for it
makes 
the currents $\bm J_{\rm L1}$ and $\bm J_{\rm L2}$ linearly independent, which
allows one to characterize the nonlinear transport by six linearly-independent parameters,  as it should be the case in systems of $C_1$ point symmetry.
We give the name `Nonlinear Diagonal current' ($\bm J_{\rm NLD}$) to the sum of two contributions 
in the last line of
Eq.~\eqref{J_transp_represent} because it is oriented at $\pm 45^\circ$ to $\bm E$
when $\bm E$ is parallel to $c$ or $a$ axes, provided $C_{\rm NLD}^{(1)}=C_{\rm NLD}^{(2)}$.
The NLD current contains only polarization-dependent terms and
has the value
\begin{equation}
J_{\rm NLD}=\abs{\bm E}^2\sqrt{\qty(C_{\rm NLD}^{(1)})^2+\qty(C_{\rm NLD}^{(2)})^2},
\end{equation}
independent of the polarization direction.
Microscopically, the NLD current is caused by the skew-scattering, BCD and side-jump microscopic mechanisms.

Our results show that the photocurrents excited along $c$ and $a$ axes have similar magnitudes, see Figs.~\ref{figS2}. This observation is in line with theoretical expressions Eqs.~\eqref{j_micro} and~\eqref{BCD}, where the lowest possible symmetry of the system $C_1$ results in the same order of magnitude for all terms. Consequently, the chiral axis of the bulk Te is not special, so,
for the currents excited by linearly polarized radiation, 
the material's chirality plays no essential role.

Our results also demonstrate that all currents, $J^c_{\rm L1, L2,0}$ reverse  their directions close to the CNP, when the effective gate voltage $U_{\rm G}$ changes sign,  see Figs.~\ref{fig3} and~\ref{figS1}(f). This agrees with the theoretical results, see Eqs.~\eqref{j_micro} and~\eqref{BCD}, where the current density is odd in the charge $q$. We note that, in the sample \#B, the current's inversion point is at small positive $U_{\rm G}$. We attribute this to the presence of both types of carriers at room temperature in our narrow-gap material, which results in simultaneous presence of the electron and hole contributions to the total current. Due to difference of microscopic parameters for the conduction and valence bands, these currents have different magnitudes, and even for somewhat smaller hole densities, the valence-band contribution exceeds the conduction-band one.
Last but not least, we note that we see no specific features when the Fermi level enters the conduction band and crosses the Weyl point.

Our data for sample \#B also show that reduction of temperature results in a drastic  enhancement  of the current magnitude,  see Fig.~\ref{figS1} ($T=300$~K) and Fig.~\ref{fig9} ($T=50$~K and $4.2$~K).
This behavior also follows from Eq.~\eqref{j_micro}
which give $j \propto \tau_{\rm tr}^2$
for $\omega\tau_{\rm tr} \leq 1$. The latter condition should be fulfilled because the room-temperature mobilities yield $\omega\tau_{\rm tr} \approx 0.1$ for electrons and $\omega\tau_{\rm tr} \approx 1$ for holes at $\omega/(2\pi)=2.54$~THz relevant for the experiment.

Finally we address variation of the current with the radiation frequency. Figure~\ref{fig3} and insets show that decrease of $\omega$ results in an increase of the current contribution's magnitudes $J^c_{\rm L1,0}$ normalized by the radiation power $P_s$. As addressed above, all  currents are caused by the Drude absorption, see Eq.~\eqref{j_micro},~\eqref{BCD}.
Therefore they are proportional to the  
high-frequency conductivity
$\sigma(\omega) = \sigma_0/(1 + \omega^2\tau_{\rm tr}^2)$. For small $\omega\tau_{\rm tr}\ll1$, the absorbance is frequency independent, and the currents are equivalent to that obtained in dc transport measurements. While, as addressed above, in our experiments at room temperature $\omega\tau_{\rm tr}$ is somewhat smaller than unity, the insets in Figs.~\ref{fig3}(a) and (b) show a clear rise of the currents $J^c_{\rm L1}/P_s$ and $J^c_{0}/P_s$ at frequency decrease. We attribute this result to the fact that the square of the THz electric field acting on carriers and causing the current in our structures, $\abs{\bm E}^2$, differs from $P_s$ by a frequency dependent factor. The reason for that is a dependence of the refractive index on frequency which is substantially enhanced in the vicinity of the reststrahlen band of tellurene. The latter is close to the frequencies used in the experiments and its position, well known for bulk Te, is not studied of 2D tellurene as yet. Thus the frequency dependence of the currents in tellurene requires a future study.

\section{Summary and Outlook} 

We studied the direct current excited by polarized terahertz radiation in tellurene structures. This current is quadratic in the electric field and belongs to the class of nonlinear electron transport effects. We observed photocurrents that are sensitive to the orientation of the electric field vector, as well as to the gate voltage. 
We attribute the THz-induced currents to mechanisms arising from the low spatial symmetry of 2D tellurene, including skew scattering, the Berry curvature dipole, and side jumps during momentum scattering. 
We established the equivalence between NLH, NLL, and NLD nonlinear transport currents and LPGE current contributions induced by THz radiation.
The above analysis shows that measuring the polarization dependence allows one to distinguish between various photocurrent contributions and helps one to conclude on the microscopic processes.
Our results demonstrate that studying polarization-dependent THz or GHz radiation-induced photocurrents paves the way for exploring the multifaceted picture of nonlinear transport in low-dimensional and bulk semiconductor materials.

\acknowledgments
The financial support of the Deutsche Forschungsgemeinschaft (DFG, German Research Foundation) via Project-IDs 564981228 (MO 5219/1-1) and 521083032 (GA 501/19) is gratefully acknowledged.  Work of L.E.G. was funded by the German Research Foundation (DFG) as part of the German Excellence Strategy -- EXC3112/1 -- 533767171 (Center for Chiral Electronics). S.D.G. is grateful for the support of by the European Union through the ERC-ADVANCED grant TERAPLASM No. 101053716. Views and opinions expressed are, however, those of the author(s) only and do not necessarily reflect those of the European Union or the European Research Council Executive Agency. Neither the European Union nor the granting authority can be held responsible for them.

\bibliography{all_lib1.bib}

\end{document}